\documentclass[preprint,12pt]{elsarticle}

\usepackage{amssymb}

\usepackage[utf8]{inputenc}

\usepackage{enumerate}
\usepackage{amsmath}

\usepackage{amssymb}
\usepackage{amsfonts}
\usepackage{bm}
\usepackage{miller}
\usepackage{graphicx}
\usepackage{textcomp}
\usepackage{nicefrac}
\usepackage{siunitx}
\usepackage[font=footnotesize,labelfont=bf]{subcaption}
\usepackage{boldline}
\usepackage{overpic}      
\usepackage{stackengine}  %
\usepackage{placeins} 
\usepackage{datetime}
\usepackage{hyperref}
\usepackage{color}
\usepackage{comment}

\usepackage{geometry}
\journal{}

\begin{document}

\begin{frontmatter}



\title{Pattern Matching Workflows for EBSD Data
Analysis:\\Quartz Chirality Mapping}
\author[AGH]{Grzegorz Cios}
\author[AGH]{Aimo Winkelmann}

\affiliation[AGH]{organization={AGH University of Krakow, \mbox{Academic Centre for Materials and Nanotechnology (ACMiN)}},
            addressline={\mbox{al.\@ A. Mickiewicza 30}}, 
            city={\mbox{30-059 Kraków}},
            country={Poland}}

\author[AGH]{Tomasz Tokarski}
\author[AGH]{Piotr Bała}

\begin{abstract}
Pattern matching approaches to electron backscatter diffraction (EBSD) in the scanning electron microscope (SEM) provide qualitatively new possibilities for the microstructural analysis of chiral non-centrosymmetric phases due to the influence of dynamical electron diffraction effects on the formation of EBSD Kikuchi patterns.
In the present study, we analyze the microstructure of polycrystalline \mbox{$\alpha$-quartz} in an agate mineral sample.
We identify characteristic intra-grain inversion domains of different handedness which are well-known from classical polarized light microscopy. 
As a result, the handedness-resolved microstructure of quartz can be imaged with the spatial and orientation resolution provided by EBSD in the SEM. 
\end{abstract}

\begin{graphicalabstract}
\includegraphics[width=\textwidth]{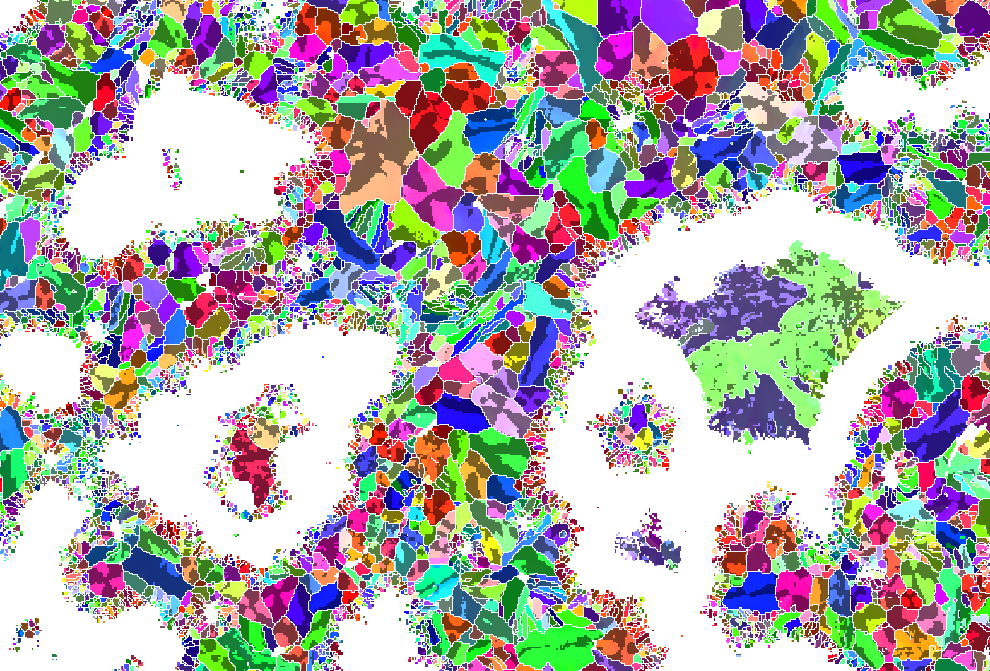}
\end{graphicalabstract}

\begin{highlights}
\item Spatial mapping of the local handedness of polycrystalline quartz samples by EBSD 
\end{highlights}


\end{frontmatter}


\section{Introduction}

Chirality is a fundamental concept that plays a crucial role across various scientific disciplines, including physics, materials science, geology, and biology
\cite{barron2009chapter,avnir2024minerals,cho2023nrb,lee2022symmetry,fecher2022materials,gregorio2020ncomm}.
While there is a range of analysis methods for the characterization of chiral polymorphs \cite{yu1998pstt}, chirality-sensitive electron diffraction in the SEM can be particularly suitable for studies of micro- and nanocrystalline materials
\cite{winkelmann2015um,winkelmann2023mc,burkhardt2021sciadv,burkhardt2020srep}.
This is because electron diffraction is sensitive to the breaking of Friedels rule by \mbox{non-centrosymmetric} crystal structures due to the strong influence of dynamical scattering \cite{nolze2015jac}.

\begin{figure}[htb!]
\begin{center}
    \includegraphics[width=8cm]{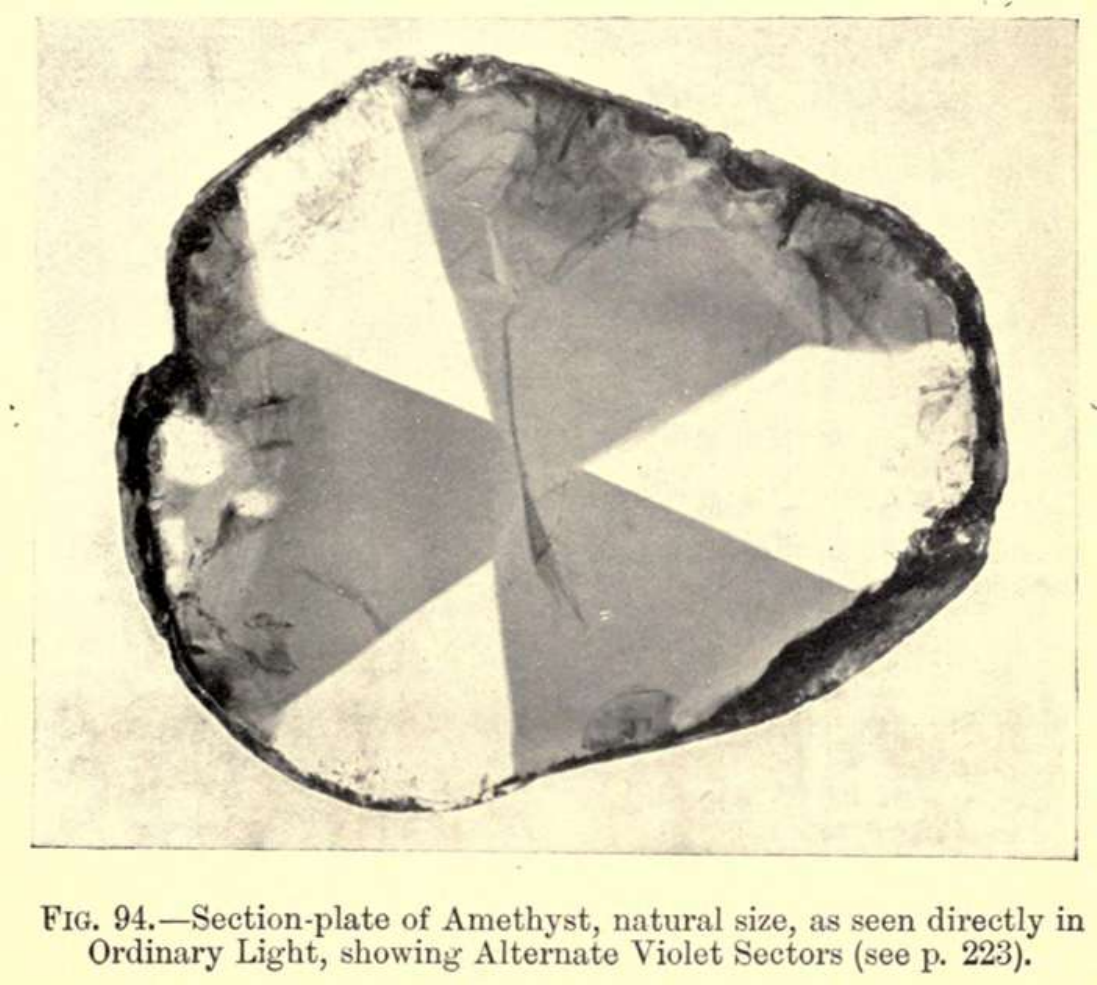}
    \vspace{-3ex}
    \end{center}
        \caption{Typical domain structure found in Amethyst, caused by domains of different handedness in quartz, reproduced from Fig. 94, page 220 in Tutton \cite{tutton1911crystals}.}
        \label{fig:tutton}
\end{figure}

In this context, trigonal $\alpha$-quartz is an important non-centrosymmetric crystal structure, which is of great importance due to its role in geological processes, as well as in industrial applications \cite{heaney1994,goetze2012}.
Concerning the effects of handedness in quartz, in Figure \ref{fig:tutton} we reproduce a classical example of an Amethyst gemstone \cite{tutton1911crystals}. 
The domain structure visible in Figure \ref{fig:tutton} is caused by twinning \cite{quartzpage_brazil}, and the dark and light domains are caused by the handedness of the underlying quartz crystal structure.

Optical characterization techniques of the handedness of quartz crystals are based on the polarization properties of light \cite{tutton1911crystals,skalwold2015,jovanovski2022ct}, but these methods lack the the spatial and orientation resolution which is possible by techniques such as electron backscatter diffraction (EBSD) in the SEM.
In various previous studies, electron diffraction techniques were applied to the characterization of quartz and provided insights into the microstructure and possible deformation mechanisms
\cite{lloyd1986jsg,lloyd1997tp,wenk2009pcm,hamers2016pcm,lloyd2024mapping,miyajima2024ejm,dey2024jsg}.
The widespread analysis of crystal handedness by EBSD, however, was not possible so far, because commercial EBSD systems were limited to centrosymmetric crystal information.

Using single Kikuchi diffraction patterns from specimens of known chirality, it was previously demonstrated that, fundamentally, EBSD can be sensitive to the handedness of quartz \cite{winkelmann2015um}.
In the present study, we extend these results by demonstrating the practical mapping of quartz chirality in a polycrystalline microstructure of an agate sample using a full pattern matching approach \cite{winkelmann2020jm,trimby2024um}. 
This extends the application of EBSD to the important problem of microstructural quartz chirality, which so far could not be studied routinely using EBSD.

\section{Experimental Details}
The EBSD setup applied in this study consists of a field emission gun SEM Versa 3D (FEI, currently Thermo Fisher Scientific) equipped with a Symmetry S2 EBSD detector (Oxford Instruments Nanoanalysis). 
The EBSD detector was operated in "Speed 1" mode with $622 \times 512$ pixel resolution and 25.4 ms acquisistion time per frame. 
The other data acquisition parameters were: "low vacuum" mode with a chamber pressure of 20 Pa, beam voltage 15 kV, beam current approx. 10 nA, sample tilt $70^{\circ}$, acquisition software Aztec 6.1 SP2 (Oxford Instruments Nanoanalysis) and export to the HDF5-based, open data format H5OINA \cite{h5oina2024} for data analysis and post-processing in AztecCrystal 3.2.
We used a sample that was cut from an agate gemstone \cite{goetze2020minerals} specimen of unspecified geographical origin.
Agates can contain intricate layered microstructures with different microcrystalline aggregates \cite{goetze2020minerals,jovanovski2022ct}, but the present investigation mainly focuses on an area containing quartz grains having sizes in the order of a few tens of $\mu$m, in order to investigate the possible presence of systematic effects of handedness in these grains.

For the pattern matching indexing and analysis using the AztecCrystal 3.2 MapSweeper software (Oxford Instruments Nanoanalysis) \cite{trimby2024um}, the crystal structure of $\alpha$-Quartz in space group $P\, 3_2\, 2\, 1$ was assumed according to the structure data published by Levien \textit{et al.} in \cite{levien1980am}, available as a CIF file in the Crystallography Open Database, with the COD code \textrm{9000775}.
The MapSweeper software provides the possibibility to either re-index acquired Kikuchi pattern data from scratch, or to refine on previously obtained results using the conventional Hough-based approach. 
The workflow can also include a self-consistent repair indexing of potentially wrongly indexed points based on the solutions of the neighboring map points \cite{trimby2024um}.
The quality of the solutions found is quantified by the normalized cross-correlation coefficient $r$, and the discrimination between different phases or between different pseudosymmetries can be quantified by differences of the respective cross-correlation coefficients.

\section{Results}

\begin{figure}[htb!]
\begin{center}
     \centering
     \begin{subfigure}[b]{0.51\textwidth}
        \centering
        \includegraphics[width=\textwidth]{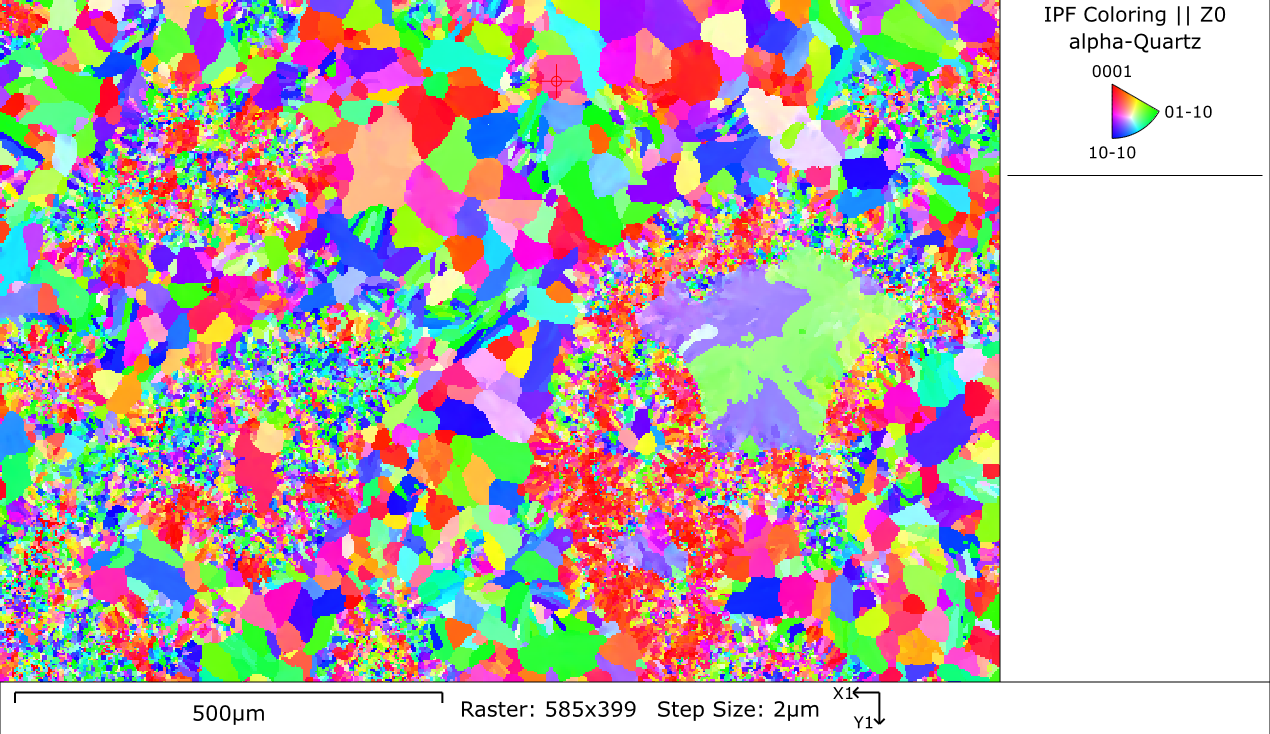}
        \caption{IPF-Z map for Quartz, Laue group $\overline{3}$m, obtained by dynamic template matching \cite{trimby2024um}}
     \end{subfigure}
     \begin{subfigure}[b]{0.51\textwidth}
         \centering
    \includegraphics[width=\textwidth]{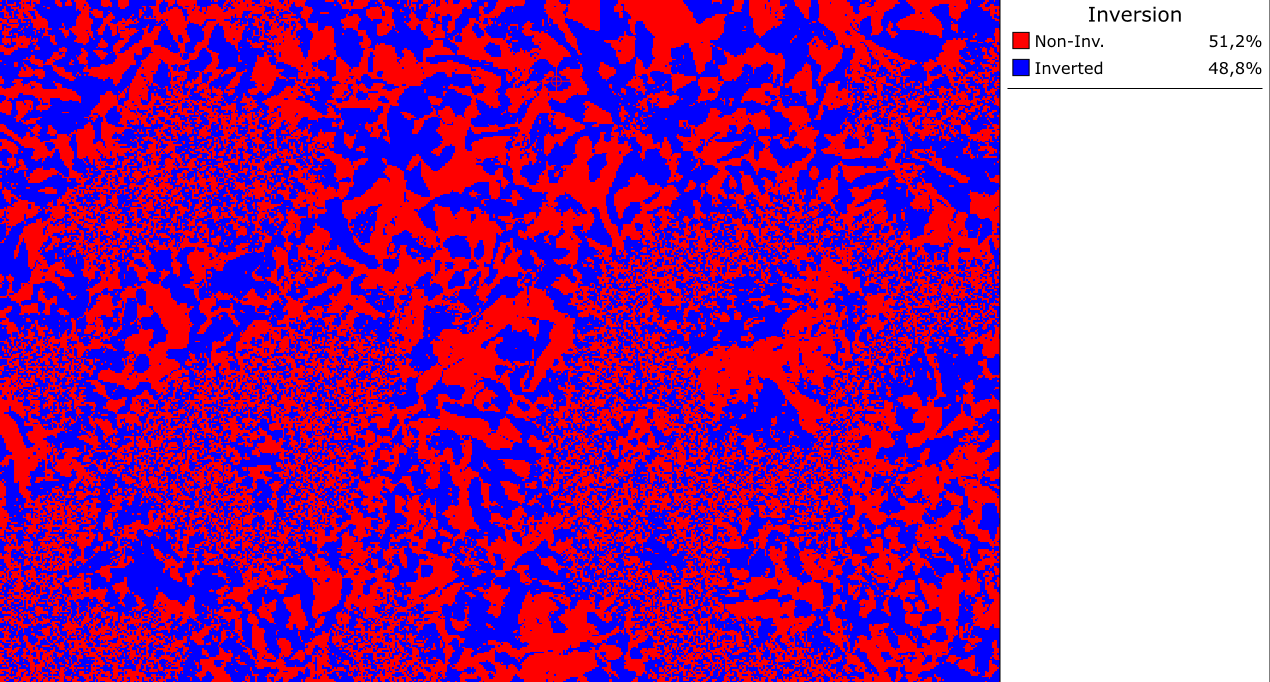}
         \caption{Inversion signal}
     \end{subfigure}
     \begin{subfigure}[b]{0.51\textwidth}
        \centering
    \includegraphics[width=\textwidth]{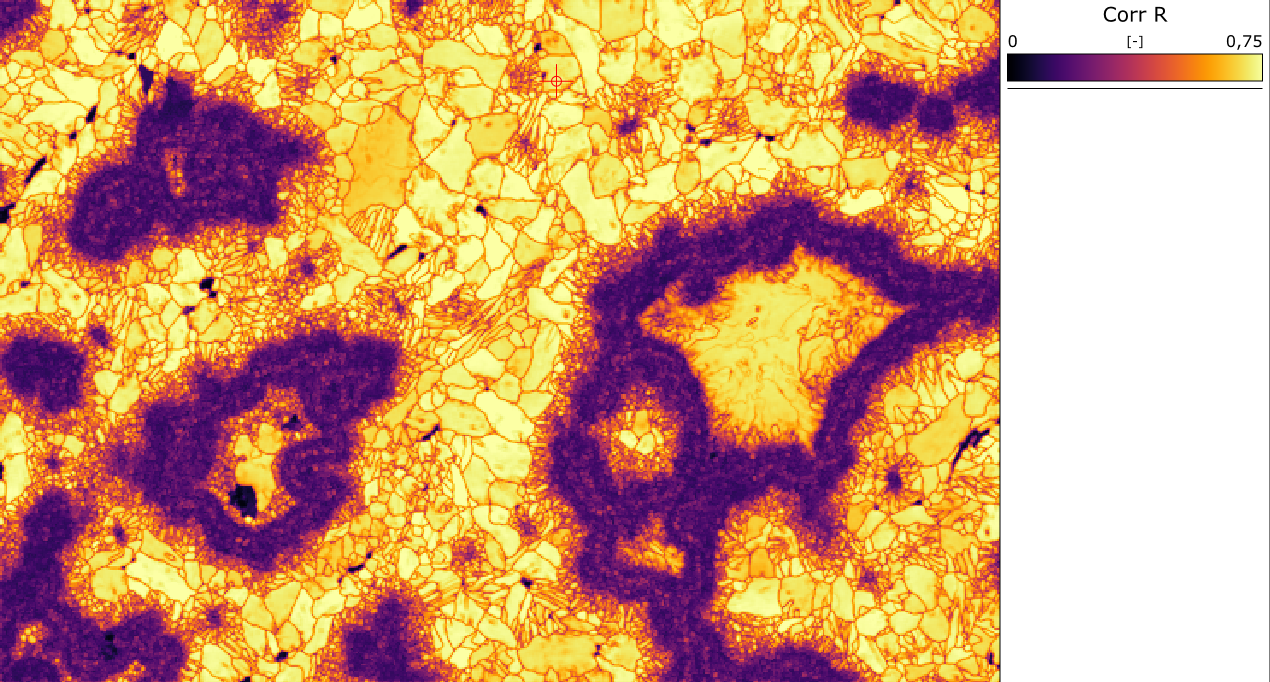}
        \caption{Normalized Cross Correlation Coefficient $r$}
     \end{subfigure}
     \begin{subfigure}[b]{0.51\textwidth}
         \centering
   \includegraphics[width=\textwidth]{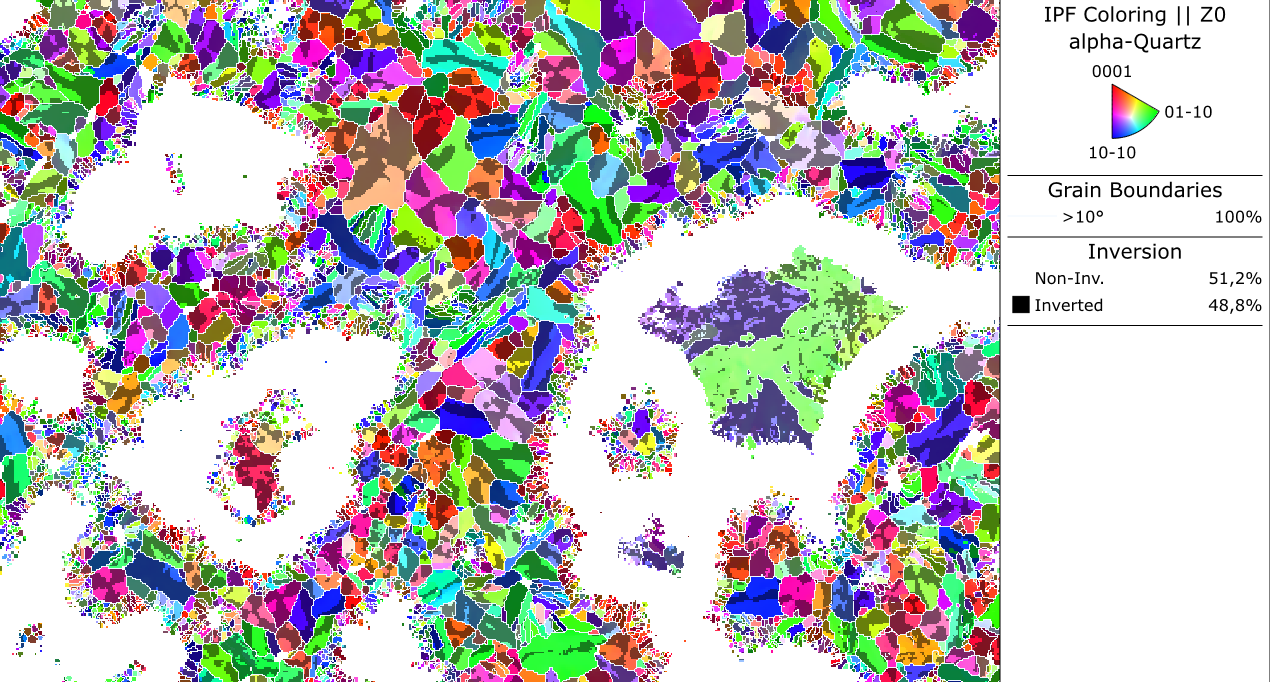}
         \caption{Combined IPF/Inversion Map, threshold $r>0.4$}
     \end{subfigure}
    \vspace{-3ex}
    \end{center}
        \caption{Results of quartz chirality EBSD mapping }
        \label{fig:maps}
\end{figure}

In Figure \ref{fig:maps}, we show results of mapping the microstructure of an agate sample. 
In the IPF map shown in Figure \ref{fig:maps}(a), the larger quartz grains have sizes in the order of a few tens of $\mu$m, while some included microcrystalline regions typical of agates 
\cite{goetze2020minerals} contain significantly smaller grains which are not the focus of the current investigation. 

The inversion signal map in Figure \ref{fig:maps}(b) has been obtained by testing for the inversion pseudosymmetry in the refinement step available in AztecCrystal 3.2 MapSweeper.
The blue color indicates a better fit of the simulated pattern of the inverted crystal structure, and we can see systematic domains in the larger grains.
The smaller grains show a more random pattern of red and blue, which we attribute to the low quality of the patterns measured from these regions, as seen by the low correlation coefficients shown in Figure \ref{fig:maps}(c).

The combined inversion and orientation information of crystal phases in EBSD can be conveyed as a point-group specific color key \cite{nolze2016jac}. 
Using an alternative approach, in Figure \ref{fig:maps} we separated the orientation and inversion information \cite{bunge1980jac} into a combination of (a) the traditional IPF map corresponding to the sample-coordinate system alignment of crystallographic axes rotated by Laue group orientations and into (b) an additional two-valued (red/blue) inversion signal map that indicates if the simulated Kikuchi pattern of the inverted or of the non-inverted structure fits better to an observed pattern, for the same Euler angles.

For visualization purposes, we overlay a partially transparent inversion signal on top of the Laue-group based IPF map in Figure \ref{fig:maps}(d).
This means, that, for an IPF-Z map, the unchanged basic IPF color for a direction \hkl[uvw] (equivalent to \hkl[-u-v-w]) in the Laue group fundamental zone on the sphere corresponds to all crystal orientations of the non-inverted reference structure with the crystallographic directions \hkl[uvw] or \hkl[-u-v-w] along the $+Z$ sample direction. 
We recall that the directions \hkl[uvw] and \hkl[-u-v-w] are indistinguishable from each other within the Laue group, which by definition is centrosymmetric.  

In order to visualize the additional inversion information without completely loosing the familiar EBSD Laue group color keys, the darker domains modify the basic Laue-group IPF color by the partially transparent inversion signal overlay and indicate that the pattern simulation for the same orientation, but using the inverted crystal structure, provides a better fit to the experimentally observed Kikuchi pattern. 
We think that this approach is especially useful to visualize domains of opposite handedness in large grains, as the relatively large change in color brightness is easily connected with a discrete change as given by the inversion signal.
In contrast to the pattern matching approach, conventional Hough-based EBSD is not sensitive to non-centrosymmetric phases, i.e. the difference of a given orientation of a crystal structure and of the same orientation (i.e. fixed Euler angles) using the inverted atomic coordinates cannot be experimentally resolved for non-centrosymmetric phases.
This is due to the fact that, effectively, a center of symmetry is introduced due to the limitations of the crystallographical analysis method which used in conventional EBSD.

\begin{figure}[htb!]
\begin{center}
    \includegraphics[width=8cm]{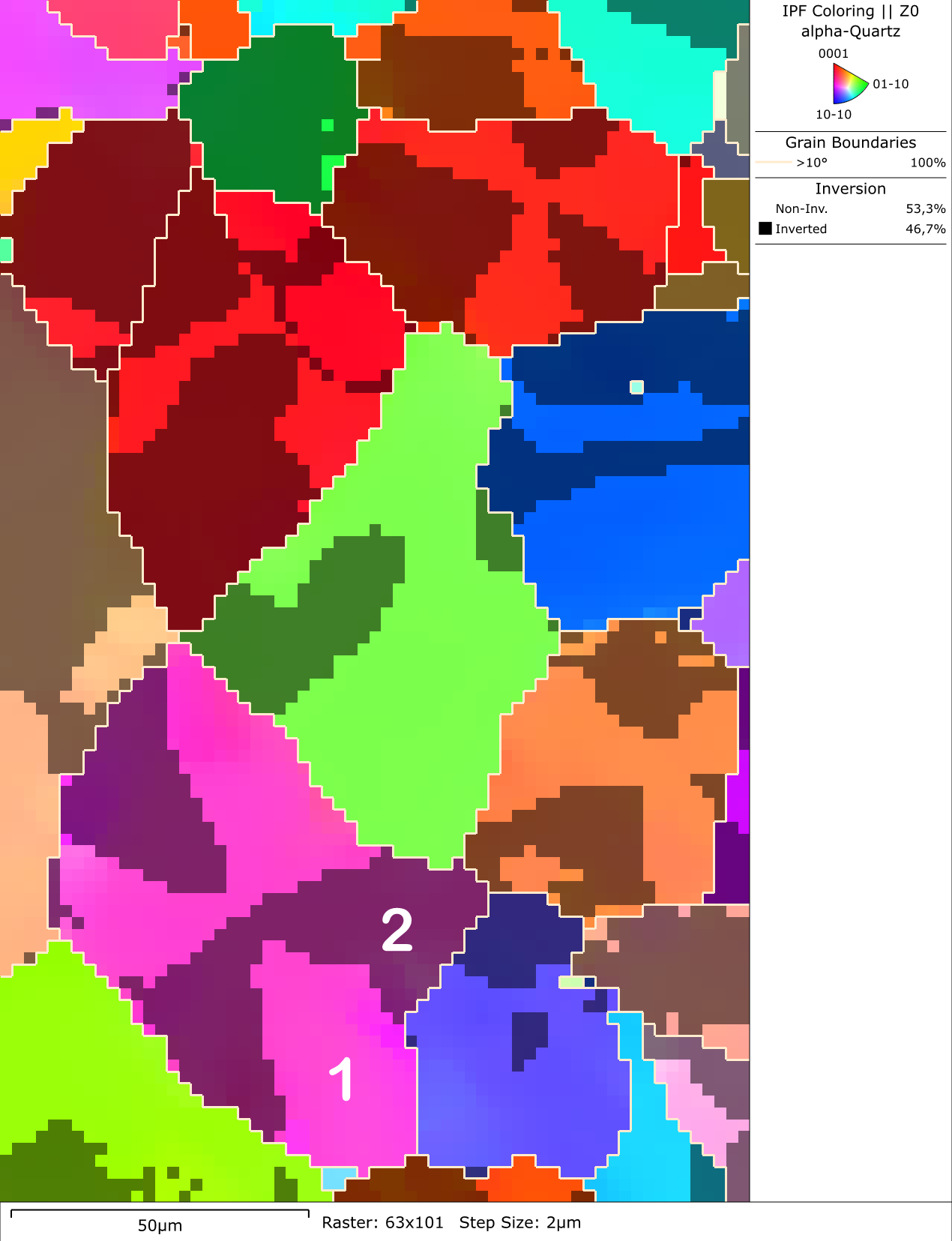}
    \vspace{-3ex}
    \end{center}
        \caption{Close-up of inversion domain structures which are characteristic for intra-grain regions of different handedness in quartz (compare Figure \ref{fig:tutton}). A pattern analysis for positions 1 and 2  is shown in Figure \ref{fig:patterns}}
        \label{fig:subset}
\end{figure}

In order to visualize only the most significant results, in Figure \ref{fig:maps}(d) we excluded all map points with $R<0.4$ which we consider as unreliable for the purpose of chirality mapping in the present study. 
For the remaining map points, we show the IPF color combined with partially transparent layer which shows the inverted structure domains by darker colors in the same IPF-color grain.
Overall, the IPF color distribution of the grains in Figure \ref{fig:maps}(a) does not seem to indicate a significant orientation texture, but almost every grain contains inversion domains.
In some of the grains, we can see the very characteristic triangular features of Brazil law twins in quartz \cite{mclaren1982pcm,quartzpage_brazil,ametrine_mindat}, for which we show a close-up in Figure \ref{fig:subset}.

\begin{figure}[htb!]
\begin{center}
     \centering
     \begin{subfigure}[b]{\textwidth}
        \centering
        \includegraphics[width=\textwidth]{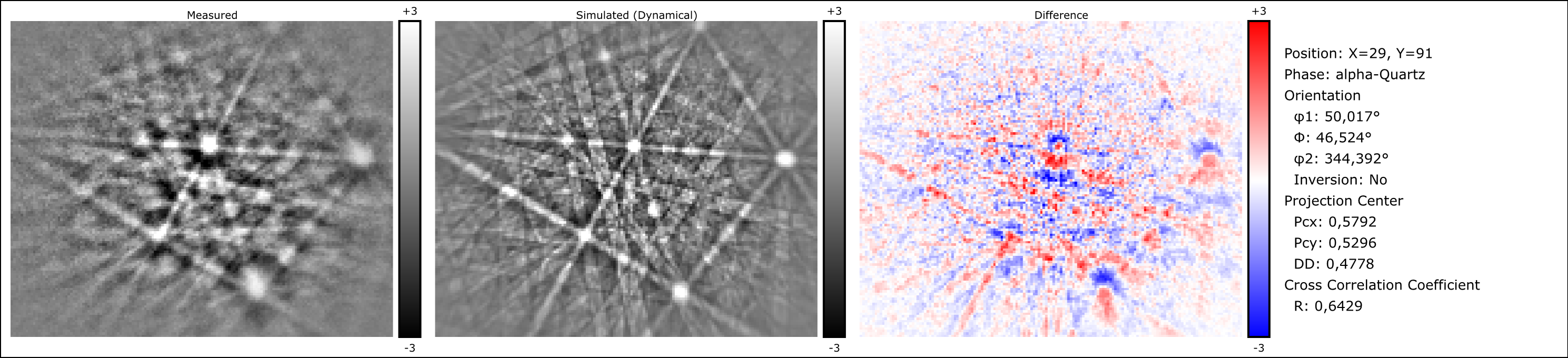}
        \caption{Position 1, test for non-inverted solution $r_{1}^{+}=0.6429$}
     \end{subfigure}
     \begin{subfigure}[b]{\textwidth}
         \centering
    \includegraphics[width=\textwidth]{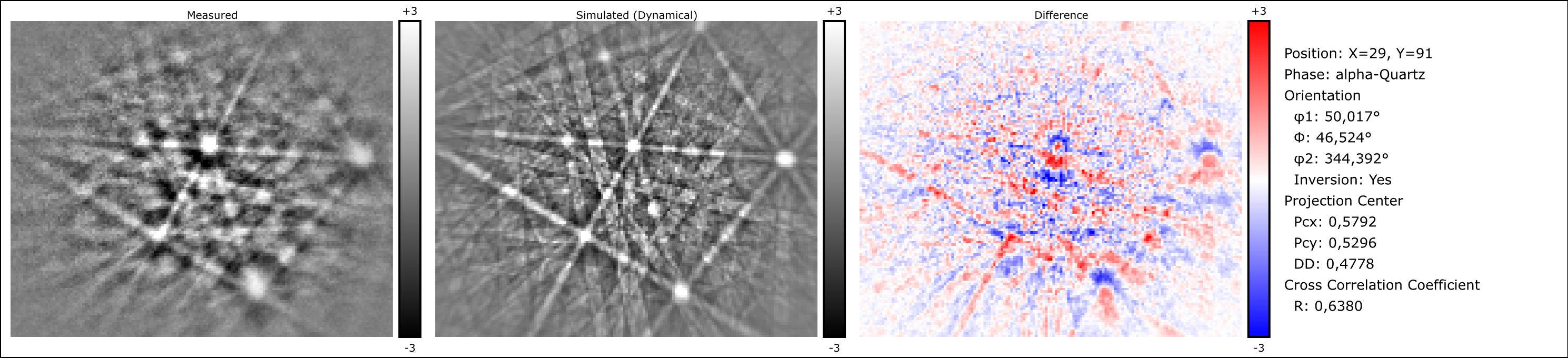}
         \caption{Position 1, test for inverted solution $r_{1}^{-}=0.6380$}
     \end{subfigure}
     \begin{subfigure}[b]{\textwidth}
        \centering
    \includegraphics[width=\textwidth]{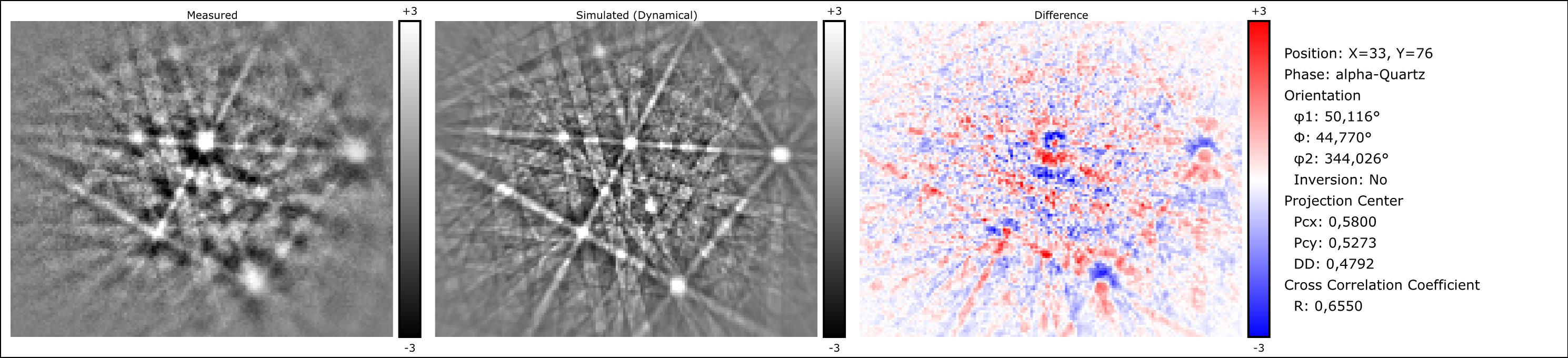}
        \caption{Position 2, test for non-inverted solution $r_{2}^{+}=0.6550$}
     \end{subfigure}
     \begin{subfigure}[b]{\textwidth}
         \centering
   \includegraphics[width=\textwidth]{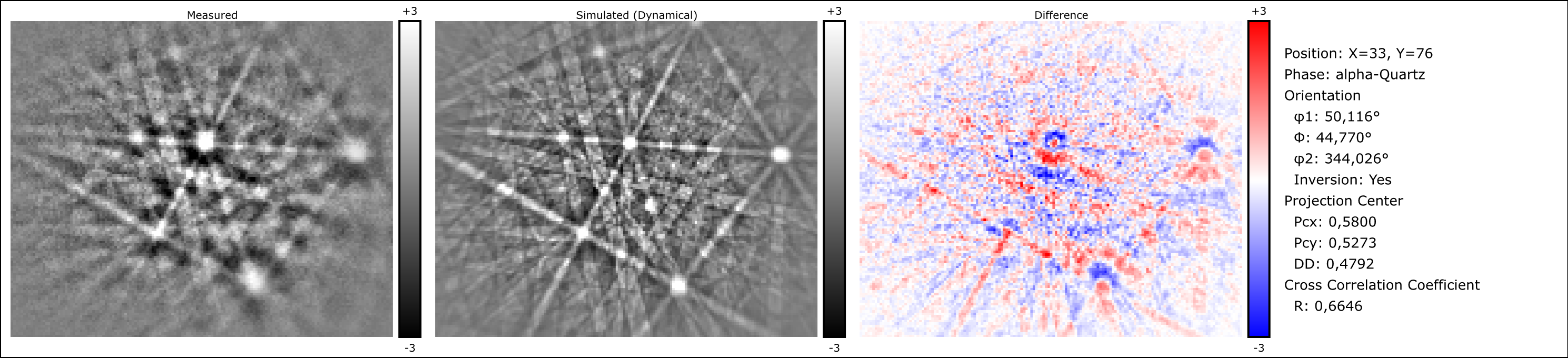}
         \caption{Position 2, test for inverted solution $r_{2}^{-}=0.6646$}
     \end{subfigure}
    \vspace{-3ex}
    \end{center}
        \caption{Examples of handedness determination from quartz Kikuchi patterns at positions 1 and 2 in the map shown in Figure \ref{fig:subset}. Pixel resolution: $155 \times 128$, supersampling: $7 \times 7$, with optimized best-fit projection center and orientation for each position, and testing the correlation coefficients for simulations of the non-inverted ($r^{+}$) and inverted structure ($r^{-}$), respectively.}
        \label{fig:patterns}
\end{figure}

In Figure\,\ref{fig:patterns}, we explicitly demonstrate the assignment of the handedness from measured Kikuchi patterns collected from domains 1 and 2 in Figure \ref{fig:subset}. 
The fit of the experimental pattern to the simulation is quantified by the normalized correlation coefficient $r$. 
For each of the two positions, we use the optimized fixed projection center and orientation, and we only switch the handedness of the simulation to test both possible crystal structures, leading to the respective correlation coefficients $r^{+}$ (crystal structure as defined in the CIF, non-inverted) and  $r^{-}$ (inverted atomic coordinates of the crystal structure defined in the CIF).

For the experimental pattern from the domain at position 1, we find  $r_{1}^{+}=0.6429$ for the simulation from the non-inverted structure, in comparison to $r_{1}^{-}=0.6380$ for the inverted structure. 
The difference $\Delta r_1 = r_{1}^{+} - r_{1}^{-}$ of the correlation coefficients is $\Delta r_1 =+0.0049$, favouring the non-inverted ($+$) reference structure.
For the experimental pattern from the domain at position 2, we find  $r_{2}^{+}=0.6550$ for the simulation from the non-inverted structure, in comparison to $r_{2}^{-}=0.6646$ for the inverted structure. This gives $\Delta r_2 = -0.0096$ which means that the inverted ($-$) reference structure is present at position 2.
The smaller absolute value of the differences for domain 1 could be caused by the presence of polysynthetic twins (e.g. layers of different handedness) \cite{mclaren1982pcm} leading to a minor contribution of the opposite handedness in region 1 within the probing value of EBSD.

We also note that we do not simply identify domains of different handedness, but we can identify the underlying crystal structure even in an absolute sense \cite{burkhardt2020srep,burkhardt2021sciadv}, i.e. we can assign the correct atomic coordinates in the unit cell of the crystal in the laboratory system. This is another advantage compared to optical methods, which usually need to calibrate the sense of the optical rotation on a reference crystal of known handedness.

\FloatBarrier
\section{Summary}

In our study, we explored the chirality of polycrystalline quartz within an agate sample using a pattern matching approach to Electron Backscatter Diffraction (EBSD). 
By leveraging the dynamical effects in Kikuchi patterns, we introduced a practical methodology for mapping quartz chirality at high spatial and orientation resolution, surpassing the limitations of traditional optical methods as well as of Hough-based EBSD techniques. 

Using an optimized hybrid analysis approach as introduced in \cite{trimby2024um}, we identified inversion domains of differing handedness within quartz grains, and we demonstrated the potential for routine chirality analysis as a tool for understanding the complex handedness behavior possible in quartz.

We also find that the simulation-based approach enables us to obtain high-resolution microstructural information with reduced requirements on the collected pattern resolution, thus optimizing the available analysis speed and reducing the data storage requirements.

\FloatBarrier
\section*{Data availability}
The analyzed data is available as H5OINA files at: \\ \url{https://dx.doi.org/10.5281/zenodo.14652785}.

\FloatBarrier
\section*{Acknowledgements}
This work was supported by the Polish National Science Centre (NCN), grant number \\ \mbox{2020/37/B/ST5/03669}.
The research results presented in this paper have been developed with the use of equipment financed from the funds of the "Excellence Initiative - Research University" program at the AGH University of Krakow. 


\begin{thebibliography}{10}
\expandafter\ifx\csname url\endcsname\relax
  \def\url#1{\texttt{#1}}\fi
\expandafter\ifx\csname urlprefix\endcsname\relax\def\urlprefix{URL }\fi
\expandafter\ifx\csname href\endcsname\relax
  \def\href#1#2{#2} \def\path#1{#1}\fi

\bibitem{barron2009chapter}
L.~D. Barron, {An Introduction to Chirality at the Nanoscale}, in: D.~B. Amabilino (Ed.), Chirality at the Nanoscale: Nanoparticles, Surfaces, Materials and more., WILEY-VCH, Weinheim, 2009, pp. 1--27.
\newblock \href {https://doi.org/10.1002/9783527625345.ch1} {\path{doi:10.1002/9783527625345.ch1}}.

\bibitem{avnir2024minerals}
D.~Avnir, {Chiral Minerals}, Minerals 14~(10) (2024).
\newblock \href {https://doi.org/10.3390/min14100995} {\path{doi:10.3390/min14100995}}.

\bibitem{cho2023nrb}
N.~H. Cho, A.~Guerrero-Martínez, J.~Ma, S.~Bals, N.~A. Kotov, L.~M. Liz-Marzán, K.~T. Nam, Bioinspired chiral inorganic nanomaterials, Nature Reviews Bioengineering 1~(2) (2023) 88--106.
\newblock \href {https://doi.org/10.1038/s44222-022-00014-4} {\path{doi:10.1038/s44222-022-00014-4}}.

\bibitem{lee2022symmetry}
C.~Lee, J.~Weber, L.~Rodriguez, R.~Sheppard, L.~Barge, E.~Berger, A.~Burton, {Chirality in Organic and Mineral Systems: A Review of Reactivity and Alteration Processes Relevant to Prebiotic Chemistry and Life Detection Missions }, Symmetry 14~(3) (2022) 460.
\newblock \href {https://doi.org/10.3390/sym14030460} {\path{doi:10.3390/sym14030460}}.

\bibitem{fecher2022materials}
G.~H. Fecher, J.~Kübler, C.~Felser, {Chirality in the Solid State: Chiral Crystal Structures in Chiral and Achiral Space Groups}, Materials 15~(17) (2022) 5812.
\newblock \href {https://doi.org/10.3390/ma15175812} {\path{doi:10.3390/ma15175812}}.

\bibitem{gregorio2020ncomm}
M.~C. di~Gregorio, L.~J.~W. Shimon, V.~Brumfeld, L.~Houben, M.~Lahav, M.~E. van~der Boom, Emergence of chirality and structural complexity in single crystals at the molecular and morphological levels, Nature Communications 11~(1) (Jan. 2020).
\newblock \href {https://doi.org/10.1038/s41467-019-13925-5} {\path{doi:10.1038/s41467-019-13925-5}}.

\bibitem{yu1998pstt}
L.~Yu, S.~M. Reutzel, G.~A. Stephenson, Physical characterization of polymorphic drugs: an integrated characterization strategy, {Pharmaceutical Science \& Technology Today} 1~(3) (1998) 118--127.
\newblock \href {https://doi.org/10.1016/s1461-5347(98)00031-5} {\path{doi:10.1016/s1461-5347(98)00031-5}}.

\bibitem{winkelmann2015um}
A.~Winkelmann, G.~Nolze, {Chirality determination of quartz crystals using Electron Backscatter Diffraction}, Ultramicroscopy 149 (2015) 58--63.
\newblock \href {https://doi.org/10.1016/j.ultramic.2014.11.013} {\path{doi:10.1016/j.ultramic.2014.11.013}}.

\bibitem{winkelmann2023mc}
A.~Winkelmann, G.~Cios, T.~Tokarski, P.~Ba{\l}a, Y.~Grin, U.~Burkhardt, {Assignment of chiral elemental crystal structures using Kikuchi diffraction}, Materials Characterization 196 (2023) 112633.
\newblock \href {https://doi.org/10.1016/j.matchar.2022.112633} {\path{doi:10.1016/j.matchar.2022.112633}}.

\bibitem{burkhardt2021sciadv}
U.~Burkhardt, A.~Winkelmann, H.~Borrmann, A.~Dumitriu, M.~König, G.~Cios, Y.~Grin, {A}ssignment of enantiomorphs for the chiral allotrope $\beta$-{Mn} by diffraction methods, Science Advances 7~(20) (2021) eabg0868.
\newblock \href {https://doi.org/10.1126/sciadv.abg0868} {\path{doi:10.1126/sciadv.abg0868}}.

\bibitem{burkhardt2020srep}
U.~Burkhardt, H.~Borrmann, P.~Moll, M.~Schmidt, Y.~Grin, A.~Winkelmann, Absolute structure from scanning electron microscopy, Scientific Reports 10 (2020) 4065.
\newblock \href {https://doi.org/10.1038/s41598-020-59854-y} {\path{doi:10.1038/s41598-020-59854-y}}.

\bibitem{nolze2015jac}
G.~Nolze, C.~Grosse, A.~Winkelmann, Kikuchi pattern analysis of noncentrosymmetric crystals, Journal of Applied Crystallography 48~(5) (2015) 1405--1419.
\newblock \href {https://doi.org/10.1107/s1600576715014016} {\path{doi:10.1107/s1600576715014016}}.

\bibitem{heaney1994}
P.~J. Heaney, C.~T. Prewitt, G.~V. Gibbs (Eds.), {Silica : Physical Behavior, Geochemistry, and Materials Applications}, 1st Edition, no. v.29 in Reviews in Mineralogy and Geochemistry Series, Mineralogical Society of America, Boston, 1994.

\bibitem{goetze2012}
J.~Götze, R.~Möckel (Eds.), {Quartz: Deposits, Mineralogy and Analytics}, Springer Berlin Heidelberg, 2012.
\newblock \href {https://doi.org/10.1007/978-3-642-22161-3} {\path{doi:10.1007/978-3-642-22161-3}}.

\bibitem{tutton1911crystals}
A.~E.~H. Tutton, \href{https://archive.org/details/crystals00tuttiala}{Crystals}, Kegan Paul, Trench, Trübner \& Co Ltd., London, 1911.
\newline\urlprefix\url{https://archive.org/details/crystals00tuttiala}

\bibitem{quartzpage_brazil}
A.~C. Akhavan, \href{http://www.quartzpage.de/crs_twins.html#brazil}{{Quartz Crystals - Twinning}} (accessed Dec 4, 2024).
\newline\urlprefix\url{http://www.quartzpage.de/crs_twins.html#brazil}

\bibitem{skalwold2015}
E.~A. Skalwold, W.~A. Bassett, \href{http://www.minsocam.org/msa/OpenAccess_publications/Skalwold_Quartz_Bullseye_on_Optical_Activity/Quartz_Bullseye_on_Optical_Activity}{{Quartz: a Bull’s Eye on Optical Activity}}, Mineralogical Society of America, Chantilly, Virginia, USA, 2015.
\newline\urlprefix\url{http://www.minsocam.org/msa/OpenAccess_publications/Skalwold_Quartz_Bullseye_on_Optical_Activity/Quartz_Bullseye_on_Optical_Activity}

\bibitem{jovanovski2022ct}
G.~Jovanovski, T.~Šijakova Ivanova, I.~Boev, B.~Boev, P.~Makreski, Intriguing minerals: quartz and its polymorphic modifications, ChemTexts 8~(3) (May 2022).
\newblock \href {https://doi.org/10.1007/s40828-022-00165-2} {\path{doi:10.1007/s40828-022-00165-2}}.

\bibitem{lloyd1986jsg}
G.~E. Lloyd, C.~C. Ferguson, A spherical electron-channelling pattern map for use in quartz petrofabric analysis, Journal of Structural Geology 8~(5) (1986) 517--526.
\newblock \href {https://doi.org/10.1016/0191-8141(86)90002-7} {\path{doi:10.1016/0191-8141(86)90002-7}}.

\bibitem{lloyd1997tp}
G.~E. Lloyd, A.~B. Farmer, D.~Mainprice, Misorientation analysis and the formation and orientation of subgrain and grain boundaries, Tectonophysics 279~(1-4) (1997) 55--78.
\newblock \href {https://doi.org/10.1016/s0040-1951(97)00115-7} {\path{doi:10.1016/s0040-1951(97)00115-7}}.

\bibitem{wenk2009pcm}
H.-R. Wenk, N.~Barton, M.~Bortolotti, S.~C. Vogel, M.~Voltolini, G.~E. Lloyd, G.~B. Gonzalez, Dauphiné twinning and texture memory in polycrystalline quartz. part 3: texture memory during phase transformation, Physics and Chemistry of Minerals 36~(10) (2009) 567--583.
\newblock \href {https://doi.org/10.1007/s00269-009-0302-6} {\path{doi:10.1007/s00269-009-0302-6}}.

\bibitem{hamers2016pcm}
M.~F. Hamers, G.~M. Pennock, M.~R. Drury, Scanning electron microscope cathodoluminescence imaging of subgrain boundaries, twins and planar deformation features in quartz, Physics and Chemistry of Minerals 44~(4) (2016) 263--275.
\newblock \href {https://doi.org/10.1007/s00269-016-0858-x} {\path{doi:10.1007/s00269-016-0858-x}}.

\bibitem{lloyd2024mapping}
G.~E. Lloyd, {Mapping intragranular microstructures in quartz: the significance of Dauphiné twinning}, Geological Society, London, Special Publications 541~(1) (2024) 299--327.
\newblock \href {https://doi.org/10.1144/sp541-2022-332} {\path{doi:10.1144/sp541-2022-332}}.

\bibitem{miyajima2024ejm}
N.~Miyajima, D.~Silva~Souza, F.~Heidelbach, Dauphiné twin in a deformed quartz: characterization by electron channelling contrast imaging and large-angle convergent-beam diffraction, European Journal of Mineralogy 36~(5) (2024) 709--719.
\newblock \href {https://doi.org/10.5194/ejm-36-709-2024} {\path{doi:10.5194/ejm-36-709-2024}}.

\bibitem{dey2024jsg}
S.~Dey, S.~Chatterjee, S.~R. Ritanjali, R.~Dobe, R.~Mukherjee, S.~Mandal, S.~Gupta, {Nanoscale visualization of high-angle misorientations in quartz-rich rocks using SEM-EBSD and Atomic Force Microscopy}, Journal of Structural Geology 183 (2024) 105146.
\newblock \href {https://doi.org/10.1016/j.jsg.2024.105146} {\path{doi:10.1016/j.jsg.2024.105146}}.

\bibitem{winkelmann2020jm}
A.~Winkelmann, B.~M. Jablon, V.~S. Tong, C.~Trager-Cowan, K.~P. Mingard, Improving {EBSD} precision by orientation refinement with full pattern matching, Journal of Microscopy 277~(2) (2020) 79--92.
\newblock \href {https://doi.org/10.1111/jmi.12870} {\path{doi:10.1111/jmi.12870}}.

\bibitem{trimby2024um}
P.~Trimby, M.~Al-Mosawi, M.~Al-Jawad, S.~Micklethwaite, Z.~Aslam, A.~Winkelmann, S.~Piazolo, {The characterisation of dental enamel using transmission Kikuchi diffraction in the scanning electron microscope combined with dynamic template matching}, Ultramicroscopy 260 (2024) 113940.
\newblock \href {https://doi.org/10.1016/j.ultramic.2024.113940} {\path{doi:10.1016/j.ultramic.2024.113940}}.

\bibitem{h5oina2024}
\href{https://github.com/oinanoanalysis/h5oina/blob/master/H5OINAFile.md}{{Oxford Instruments NanoAnalysis HDF5 File Specification}}, accessed Nov. 23, 2024 (2024).
\newline\urlprefix\url{https://github.com/oinanoanalysis/h5oina/blob/master/H5OINAFile.md}

\bibitem{goetze2020minerals}
J.~Götze, R.~Möckel, Y.~Pan, {Mineralogy, Geochemistry and Genesis of Agate - A Review}, Minerals 10~(11) (2020) 1037.
\newblock \href {https://doi.org/10.3390/min10111037} {\path{doi:10.3390/min10111037}}.

\bibitem{levien1980am}
L.~Levien, C.~T. Prewitt, D.~J. Weidner, \href{https://pubs.geoscienceworld.org/msa/ammin/article-abstract/65/9-10/920/41195/Structure-and-elastic-properties-of-quartz-at}{Structure and elastic properties of quartz at pressure}, {American Mineralogist} 65~(9-10) (1980) 920--930.
\newline\urlprefix\url{https://pubs.geoscienceworld.org/msa/ammin/article-abstract/65/9-10/920/41195/Structure-and-elastic-properties-of-quartz-at}

\bibitem{nolze2016jac}
G.~Nolze, R.~Hielscher, Orientations -- perfectly colored, J. Appl. Cryst. 49 (2016) 1786--1802.
\newblock \href {https://doi.org/10.1107/S1600576716012942} {\path{doi:10.1107/S1600576716012942}}.

\bibitem{bunge1980jac}
H.~J. Bunge, C.~Esling, J.~Muller, The role of the inversion centre in texture analysis, Journal of Applied Crystallography 13~(6) (1980) 544--554.
\newblock \href {https://doi.org/10.1107/s0021889880012757} {\path{doi:10.1107/s0021889880012757}}.

\bibitem{mclaren1982pcm}
A.~C. McLaren, D.~R. Pitkethly, The twinning microstructure and growth of amethyst quartz, Physics and Chemistry of Minerals 8~(3) (1982) 128--135.
\newblock \href {https://doi.org/10.1007/bf00311283} {\path{doi:10.1007/bf00311283}}.

\bibitem{ametrine_mindat}
Mindat.org, \href{https://web.archive.org/web/20230116172356/https://www.mindat.org/min-7606.html}{Ametrine} (2023).
\newline\urlprefix\url{https://web.archive.org/web/20230116172356/https://www.mindat.org/min-7606.html}

\end{thebibliography}

\end{document}